# Cross-Domain Segmentation with Adversarial Loss and Covariate Shift for Biomedical Imaging

Bora Baydar, Savas Ozkan, A. Emre Kavur, N. Sinem Gezer, M. Alper Selver, Gozde Bozdagi Akar,

*Abstract*—Despite the widespread use of deep learning methods for semantic segmentation of images that are acquired from a single source, clinicians often use multi-domain data for a detailed analysis. For instance, CT and MRI have advantages over each other in terms of imaging quality, artifacts, and output characteristics that lead to differential diagnosis. The capacity of current segmentation techniques is only allow to work for an individual domain due to their differences. However, the models that are capable of working on all modalities are essentially needed for a complete solution. Furthermore, robustness is drastically affected by the number of samples in the training step, especially for deep learning models. Hence, there is a necessity that all available data regardless of data domain should be used for reliable methods. For this purpose, this manuscript aims to implement a novel model that can learn robust representations from cross-domain data by encapsulating distinct and shared patterns from different modalities. Precisely, covariate shift property is retained with structural modification and adversarial loss where sparse and rich representations are obtained. Hence, a single parameter set is used to perform cross-domain segmentation task. The superiority of the proposed method is that no information related to modalities are provided in either training or inference phase. The tests on CT and MRI liver data acquired in routine clinical workflows show that the proposed model outperforms all other baseline with a large margin. Experiments are also conducted on Covid-19 dataset that it consists of CT data where significant intra-class visual differences are observed. Similarly, the proposed method achieves the best performance.

## I. INTRODUCTION

Computer-aided machine learning techniques have been used frequently over the past decade to reduce the time that experts spend on the diagnoses of the patients and to reduce the failure / misinterpretation due to the human factors.

In particular, semantic segmentation can be identified as one of the emerging fields for biomedical imaging. The objective of this problem is to automatically extract semantically correlated segments of body organs for diagnosis, treatment planning and follow up.

In the literature, several promising techniques are proposed in this context. With the advance of computing units and increasing the accessibility of data, use of fully automated methods is exponentially increased. To this end, Deep Convolution Neural Networks (DCNs) become one of the most powerful techniques for semantic segmentation problem [1, 2, 5, 6, 7]. However, the methods based on DCNs are sensitive to small variations on data such as geometric and brightness. Therefore, the performance degradation become exponential for multi-domain data [19]. In addition, the specification of conventional structures hampers the performance for cross-domain segmentation that we will present further explanations in the manuscript. Note that models must compute distinct patterns from each domain without priors and they must be capable of learning robust representations for a complete solution. Otherwise, these solutions become nothing but a manual selection of models for each domain or sensor type.

In this regard, multi-domain segmentation aims to extract the organs of interest from different image series without explicitly knowing their sources (i.e. CT or MR or data with significant intra-class differences). In particular, Stein's paradox claims that there is an accurate solution (on average) when three or more parameter sets are estimated simultaneously instead of separate estimators of individual parameters. This paradox can be seen as the fundamental motivation behind the cross-domain learning.

To deploy such models, all available images must be used in the training step without knowing the acquisition device or type. Ultimately, this needs a novel approach to extract rich representations that are shared in different modalities. Hence, such models can contain shared and distinct features to segment organs in a single model.

In this manuscript, we tackle cross-domain learning for multi-domain liver and Covid-19 (i.e., lung and covid-19) segmentation problems. Our proposal is based on a modified DCN architecture (i.e. U-Net architecture) that has a hard parameter sharing for multi-domain data (i.e., all learnable parameters are shared.). In particular, we replace the order of the normalization module and convolution kernels to have covariate shift property [8] that is intentionally overlooked in conventional techniques. However, this modification is still underestimated the solution. Hence, adversarial loss is adapted to improve the predictions by increasing selectivity of responses. Precisely, kernels take more reliable statistical information into account from each individual domain so that

This work is supported in part by TUBITAK under project 116E133.

Bora Baydar, Savas Ozkan and Gozde Bozdagi Akar are with Middle East Technical University, Department of Electrical and Electronics Engineering, 06800, Ankara, Turkey (e-mail: ozkan.savas@metu.edu.tr).

A. Emre Kavur and Alper Selver are with Dokuz Eylul University, Department of Electrical and Electronics Engineering, 35390, Izmir, Turkey.

N. Sinem Gezer is with Dokuz Eylul University, Faculty of Medicine, Department of Radiology, 35330, Izmir, Turkey.

input data is clustered based on their modalities in an unsupervised manner. To this end, this improves the generalization by using the domain information contained in the training samples. Last but not least, on the contrary to the baselines, the proposed method uses data from various modalities as input without prior knowledge and all parameters are shared at the end. Notice that this solution does not primarily require modality information in both training and inference steps. For this reason, it enables us to deploy a complete model that can work on all modalities and improves the usability of the solution.

The remainder of the manuscript is organized as follows: First, we will provide a literature survey related to the basis of the proposed model and existing solutions for multi-domain. Then, the proposed method will be described in detail. Lastly, experimental results and their evaluations will be explained together with their quantitative evaluations and conclusions will be drawn.

## II. RELATED WORK

There are success studies on biomedical data segmentation in the literature. In particular, the studies often focus on semi-automatic and fully automated segmentation strategies and the objectives of which are to accelerate the diagnosis of disease by reducing operating costs. Therefore, in this section we summarize the recent semantic segmentation studies. Although the literature for multi-domain approaches is limited, baseline algorithms are explained in detail. To ease the understandability, this section is divided into two parts: semantic segmentation and cross-domain learning.

### A. Semantic Segmentation

The advantage of semi-automated methods is that these methods do not need a large collection of labeled data as in fully automated methods. In fact, human operators actively interact with data to fill semantic gaps in these techniques. The semantic information for these techniques is notelessly based on hand-crafted features computed from data by exploiting color and geometric features. To this end, performance varies with the precision of annotations (i.e., manual labels) [3]

To minimize the semantic gaps and dependence on supervised labels, fully-automated models use a large body of labeled data to provide more generalized solutions to the problem. Likewise, hand-crafted color and geometric features are extracted from data and various classification techniques can be adapted to yield superior performance. Note that the drawback of hand-crafted features is that observations are based on human interpretations about data and latent data patterns can be misused [4]. As a remedy, neural networks have recently been an essential and self-evident part of collectively distillations of patterns from data since the feature extraction and classification steps are optimized concurrently. Therefore, rich representations can be estimated [5, 6, 7] by projecting high-dimensional data to low-dimensional spaces with non-linear transformations.

In the literature, various network architectures have been used for image segmentation problem namely FCN-8 [5], U-Net [6] and deepMedic [7]. Each architecture has a unique network specification to increase the convergence rate of learnable parameters, to obtain more accurate results or to compute robust representations on low-power devices. For instance, U-Net utilizes skip connections (i.e., residual connection) to transfer the backpropagation error effectively to the initial levels. Similarly, FCN-8 fuses latent features computed from different layers improve performance while increasing computation time. Moreover, different modules can be added to these baseline architectures for better performance [18, 30].

Indeed, semantic segmentation is also applied for biomedical data [20, 21, 22].

### B. Cross-Domain Learning

Unpaired cross-domain learning face several additional challenges compared to paired conventional methods [17]. Even if the cross-domain image analysis is a very impactful direction, these challenges are rarely discussed and analyzed in the literature [25]. The most straightforward solution is to extract features about data from a dedicated model for different modalities and then fuse these features at some points to predict the segmentation [17]. Moreover, [31] hierarchically clusters data into a set of sub-surfaces based on both geometric and appearance features. Then, segments are estimated with a learning method. Notice that domain statistics are preserved in the solution in an unsupervised manner and our contribution shares similar observations in this framework. Instead of independent clustering technique, we integrate adversarial loss that can be trained with an end-to-end manner.

Recently, this idea is transferred with normalization layers where it aggressively normalizes domain-specific statistics with a domain-specific parameter set [23, 24]. By this way, instead of unimodal distributions, multi-modal distributions are fitted by domain specific normalization layers. However, the main drawback is that supervision about the modality must be provided in the parameter learning steps. Eventually, the usability of the solution is reduced.

Moreover, the studies focus on the domain adaptation by leveraging adversarial loss in the learning step. Basically, the cycle-transformation is computed between cross-domains [29, 32, 33]. However, learning such adversarial loss can lead instability due to the nature of this learning scheme [34]. Similarly, supervision is essential in the models in course of data conditions.

## III. PROPOSED METHOD

Our primary goal is to develop a novel model that can be trained with different samples from multiple imaging domains and still work well when any sample from different modalities is given as input. The superiority is that no information/prior related to the modalities of data is provided in the training and inference steps.

In this regard, we consider an input sample $x_d \in X_d$ and a collection of label spaces $Y_d$ per input domain $d$ such that input labels $\{y_d^0 \; y_d^1 \ldots, y_d^C\}$ are given where $C$ indicates the number of classes. Furthermore, we consider a single hypothesis class that covers all multiple domains as $f(x_d; \theta): X_d \to Y_d$ with a shared parameter set $\theta$. We also consider the following formulation to minimize a loss function for each input data from multiple domains:

$$\min_\theta \sum_{d=1}^{D} \alpha_d \cdot L(x_d; \theta) \qquad (1)$$

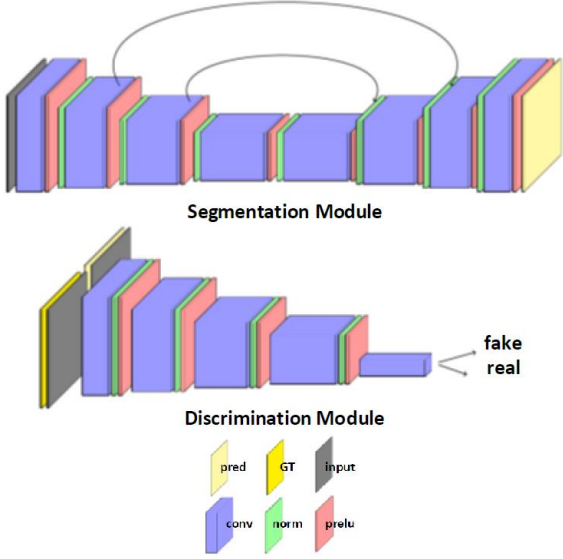

Fig. 1. The flow of the proposed method. In the propose method, two NN modules are utilized for cross-domain learning problem. Notice that no information for the modality of data is provided in both modules.

Here, $\alpha_d$ indicates the static weight term and in the scope of the manuscript, they are set to 1.

For this purpose, we exploit the framework of Deep Convolutional Neural Network (DCN) that reserves many learnable parameters for the solution. In short, this type of models consists of encoder-network modules where each of the modules is specialized for different purposes. For instance, encoder modules take an input sample and they iteratively compute abstract representations from data at each level. To this end, the dimensionality of data is reduced while the representation capacity is hierarchically increased. Later, these latent representations are aggregated at each subsequent level with the decoder modules and the element-wise summation of skip-connection is illustrated in Fig. 1.

As seen, U-Net [6] architecture is used and this architecture ultimately improves performance because the prediction error estimated from the top of networks is propagated effectively through encoder layers for better parameter convergence. Moreover, it allows us to learn incremental representations due to the iterative feature extraction steps from residual connections [2].

The segmentation output $\widehat{y_d^c}$ for an input sample $x_d$ is obtained by maximizing a true prediction loss (i.e. softmax cross-entropy loss) as follows:

$$L_{cls} = \frac{1}{C} \sum_{c=1}^{C} y_d^c \cdot \log \widehat{y_d^c} \qquad (2)$$

As expected, this loss function enforces the parameters to distinguish class segmentations by maximizing true prediction outputs while suppressing the rest. Note that other loss functions are existed in the literature [35]. But the main scope of the manuscript is to prove the effectiveness of the proposed method for cross-domain learning.

In the following sections, we first present our contributions to the structure of the DCN architecture and loss function for cross-domain learning. Precisely, normalization layer is reordered to contain domain-specific statistics with no additional parameter set or supervised label. However, we observe that this also introduces some drawbacks since the sparsity of responses (i.e., the confidence of responses) is drastically affected. For this purpose, an additional loss function is adapted to cluster visual patterns to improve the effectiveness for multi-domain distributions. In the experiments, we show that the final architecture yields state-of-the-art performance for cross-domain problem. Lastly, the data preparation and post-processing steps used in the model are also described.

*A. Cross-Domain Learning*

In the literature, convolution kernels are often used with additional layers to boost the performance of DCN models. Precisely, normalization and activation layers are integrated to the output of convolution kernels so as to regularize the distributions of kernels by reducing covariate shift as well as to enhance the non-linearity of parameters [8, 11].

With all of the above in mind, the main objective of normalization layers is to reduce the covariate shift for different distribution characteristic so that even if the content of data is changed, the conditional distribution of outputs is still unchanged. By this fact, the goal of normalization layers is to transform responses into a common distribution by highlighting only common patterns exhibited from data.

Although this elimination of covariate shift with the normalization layers increase performance for conventional problems [8], this can be harmful for cross-domain data especially by learning distinct patterns from both domains. Strictly speaking, for multi-domain data, the objective is to contain both shared and distinct patterns in the prediction step by incorporating a single model at the end.

Mathematically, convolution operation and batch normalization for an input $z$ can be simplified as $f_{conv}(z) = Wz$ (i.e. $W$ is trainable parameters) and $f_{bn}(z) = \frac{z-\mu(z)}{\sigma^2(z)}$ respectively where $\mu(.)$ and $\sigma^2(.)$ indicate the mean and variance of input samples. Note that he latent feature $z$ indicates the batch outputs of the DCN model from multi-domain data. The function $f_{post}(.)$ can be written when two operations are sequentially applied:

$$f_{post}(z) = \frac{Wz - \mu(Wz)}{\sigma^2(Wz)}, \qquad (3)$$

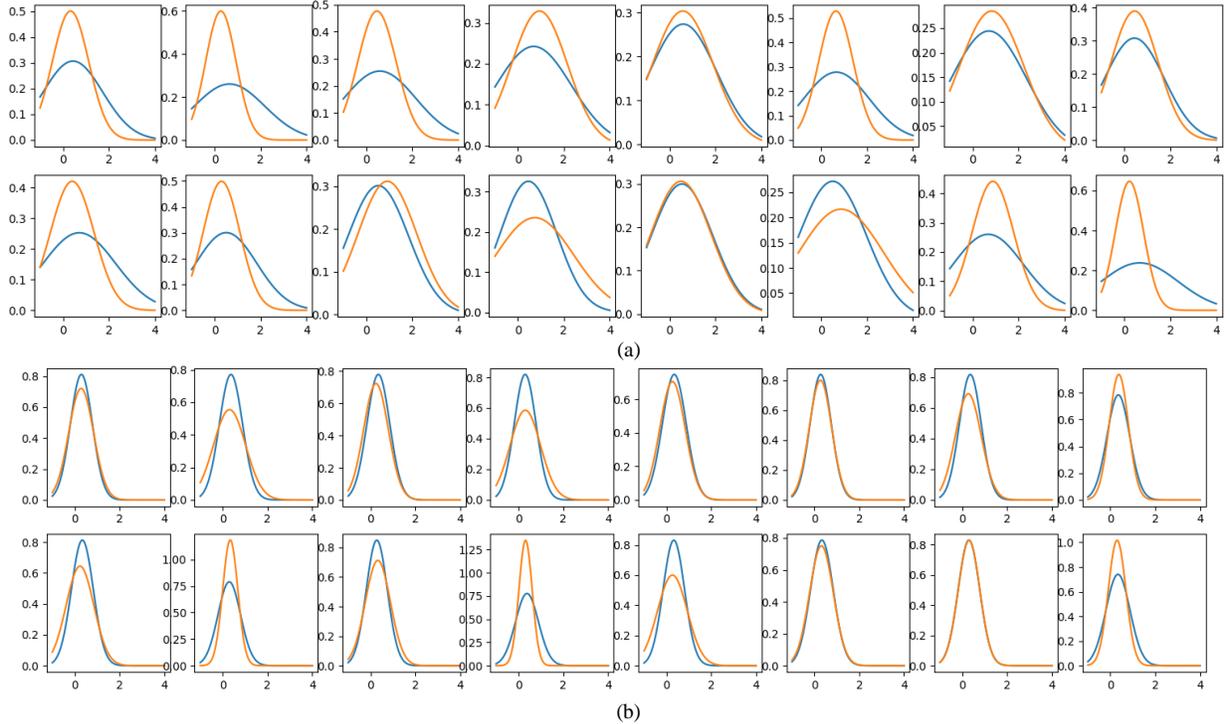

Fig. 2. The distributions of responses for CT (blue) and MR (yellow) from randomly sampled convolution kernels. (a) First two rows correspond to the distributions for the use of normalization step before convolution kernels (proposed structure). (b) Last two rows illustrate the distributions for the use of convolution kernel before normalization step (conventional structure). It is clear that the distributions for the conventional structure have similar statistical characteristics (i.e, mean and variance) for each modality, yet the response distributions for CT and MR data are more discriminative for the proposed structure. The reason is that the proposed structure takes covariate shift into account and clusters the responses according to their modalities in an unsupervised manner.

Note that these mean and variance operations are equal to $\mu(Wz) = \mathrm{E}[Wz]$ and $\sigma^2(Wz) = \mathrm{E}[(Wz - \mathrm{E}[Wz])^2]$ respectively. As noticed, statistical estimates (i.e., mean and variance) depend on the trainable parameter $W$. In this case, covariate shift is suppressed by normalizing the input feature space where only common data patterns are kept after the transformation. However, as mentioned, this is not desirable for cross-domain data since distinct features for each data type is lost as the function characteristic.

Therefore the order of the normalization layer and convolution kernels is replaced so that the effect of affine transformations to the normalization step is partially reduced. Ultimately, more statistical information about each modality is taken into account before the transformation and this increases the distinctiveness of responses for each domain and the formulation for the pre-structure $f_{pre}(.)$ is updated as follows:

$$f_{pre}(z) = \frac{Wz - W\mu(z)}{\sigma^2(z)}, \quad (4)$$

In this case, mean and variance are equal to $\mu(z) = E[z]$ and $\sigma^2(z) = \mathrm{E}[(z - E[z])^2]$ respectively and depend only the distributions of input samples. A unimodal distribution from different modalities (i.e., MR, CT or intra-class data distributions) are calculated at each level before the affine transformations.

Note that we still fit a unimodal distribution instead of a multimodal distribution for multi-domain data (i.e., a single mean and variance parameters) in the normalization step [23, 24]. Use of this model alone eventually underestimates the solution even if it is partially improved. Hence, we incorporate adversarial model to reproduce the true distributions from multi-domain data.

The initial application of adversarial models for semantic segmentation problem is to enhance the predictions at the edges of segments [5, 6, 7, 30]. However, in our method, we exploit this model to cluster the latent representations obtained from different modalities.

Note that use of adversarial loss for clustering is well studied in the literature and relatively better performance is achieved [37, 38] compared to other baselines. Another advantage is that the loss function allows us to learn parameters in an end-to-end manner.

For this purpose, we integrate a learnable framework of Conditional Adversarial Networks (CAN) to our loss function. The parameters of the model are trained in an end-to-end learning manner so that it clusters visual similarities of cross-domain data in representation space. Hence, the weakness of unimodal distribution is somehow mitigated and enhances the representation capacity.

In CAN structure, labels and predictions related to data are conditioned with input data. For instance, input data $x_d$ is conditioned with the corresponding label $y_d$ (i.e., channel-wise concatenation) and it is denoted as $u_d$. Similarly, input data $x_d$ and the prediction of the model $\hat{y}_d$ are conditioned as $\hat{u}_d$ with the same manner. To this end, an adversarial loss is computed to distinguish the real data $u_d$ and prediction $\hat{u}_d$ by learning a discriminative NN model $Disc(.)$ as:

$$L_{Disc} = \mathrm{E}[Disc(u_d)] + \mathrm{E}[1 - Disc(\hat{u}_d)] \quad (5)$$
$$L_{Gen} = \mathrm{E}[Disc(\hat{u}_d)] \quad (6)$$

where the discriminator loss $L_{Disc}$ aims to discriminate real labels from predicted labels while the generator loss $L_{Gen}$ tries

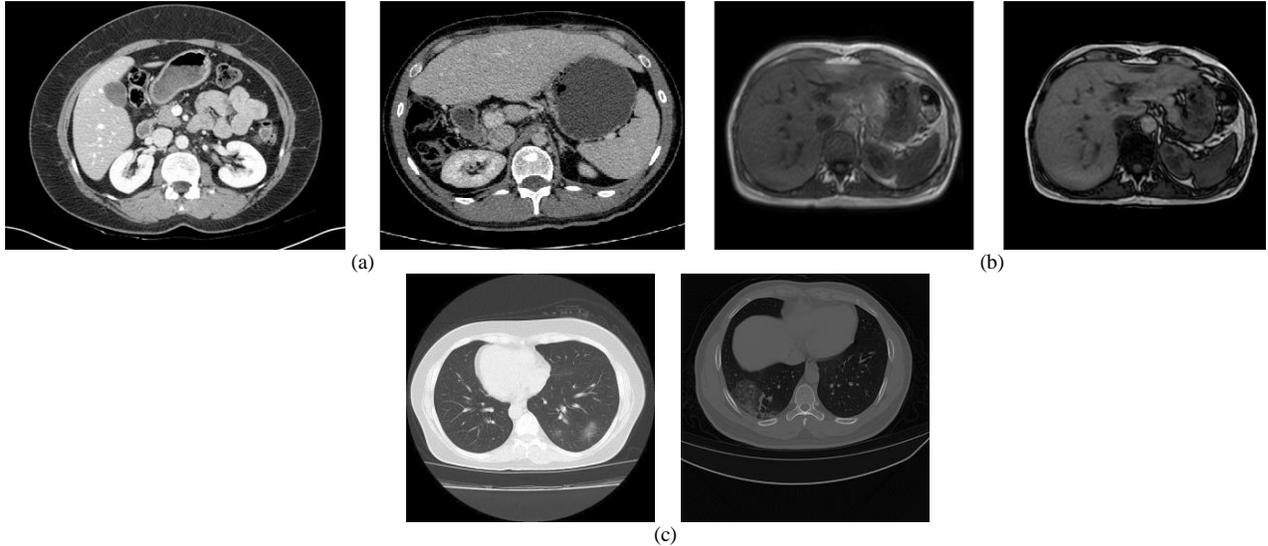
Figure 3. Sample images from Liver and Covid datasets. (a) Liver CT, (b) Liver T1-MR and (c) Covid-19 CT.

to tweak the parameters to obtain segments similar to true labels.

For this purpose, the overall loss function $L$ is equal to the combination of $L_{Gen}$ and $L_{Cls}$ by rescaling these losses as follows:

$$L = L_{Cls} + 0.001 \times L_{Gen} \qquad (7)$$

Note that the coefficient of adversarial loss $L_{Gen}$ is relatively small in comparison to prediction loss $L_{cls}$ since the main objective of the model is to predict cross-domain segmentation successfully.

To optimize the learnable parameters, Adam solver is utilized with the same configurations as in the original paper [16]. Moreover, learning rate and iteration number are set to $0.001$ and $100K$ respectively in the experiments.

To show the impact, Fig. 2 illustrates the distributions of responses for CT and MR data sampled from various kernels trained with conventional (post) and proposed (pre) structures. As seen, responses for CT and MR data behave differently for the proposed structure even if a unimodal is used. Precisely, different kernels obtain different response characteristics for each modality. The reason is that the normalization step in the proposed structure and use of adversarial loss help to contain covariate shift in the solution by clustering the responses. Hence, it enables to extract shared and distinct features from data simultaneously. Lastly, parametric Relu is selected as an activation function which rescales negative responses with a learnable parameter [11].

To show the effectiveness of the architecture, other variants of batch normalization such as instance normalization [9] are also analyzed in the manuscript. Simply, instead of batch statistics of data, spatial and kernel statistics of inputs are exploited in this variant. Experiments show that the proposed solution is still applicable to different normalization modules and the solution does not overfit to a particular architecture.

### B. Data Preparation and Post-Processing Step

The proposed solution is compatible to estimate segmentation results by accepting data with different channels. In this way, more information about the data can be supplied to the model, unless a temporal correlation between layers does not change significantly for the modalities. Formally, a sample $x_d$ corresponds to channel-wise concatenation of $T$ upper and $T$ lower neighborhood slices so that $S = 2 \times T + 1$. Note that since $T$ is a hyper-parameter that should be tuned based on the performance of modalities.

Finally, as a pre-processing step, the contrasts of pixels are rescaled to a particular range to compensate changes due to the variations in sessions and use of different imaging devices especially non-calibrated intensity ranges for MR data. To maintain the cardinality of predictions, we apply a 3D connected-component approach to eliminate small outliers that do not have connections with the largest volume of segmentation predictions (only for CT and MR data).

## IV. EXPERIMENTS

In this section, we will first describe the details of datasets, evaluation metrics and baseline techniques used in the manuscript. Later, we will mention experimental results conducted in the manuscript. These results are extensively compared with baseline techniques. Moreover, the outputs of the proposed model with a large hyper-parametric space and structural differences (i.e., it can be a modification in the architecture, presentation of data or optimization step) are examined in detail by presenting reasonable explanations.

### A. Datasets

As stated, liver and Covid datasets are utilized. First, Covid-19 dataset contains 20 labeled COVID-19 CT lung scans. Lungs and infections are labeled by radiologists. 5 scans of the dataset are reserved for the test and the rest is used in the training step. Note that this dataset exhibits large intra-class variations due to scanner type differences[1]. Visual samples from this dataset are illustrated in Fig. 3.

The second database includes unpaired CT and MRI data,

---

[1] http://medicalsegmentation.com/covid19/

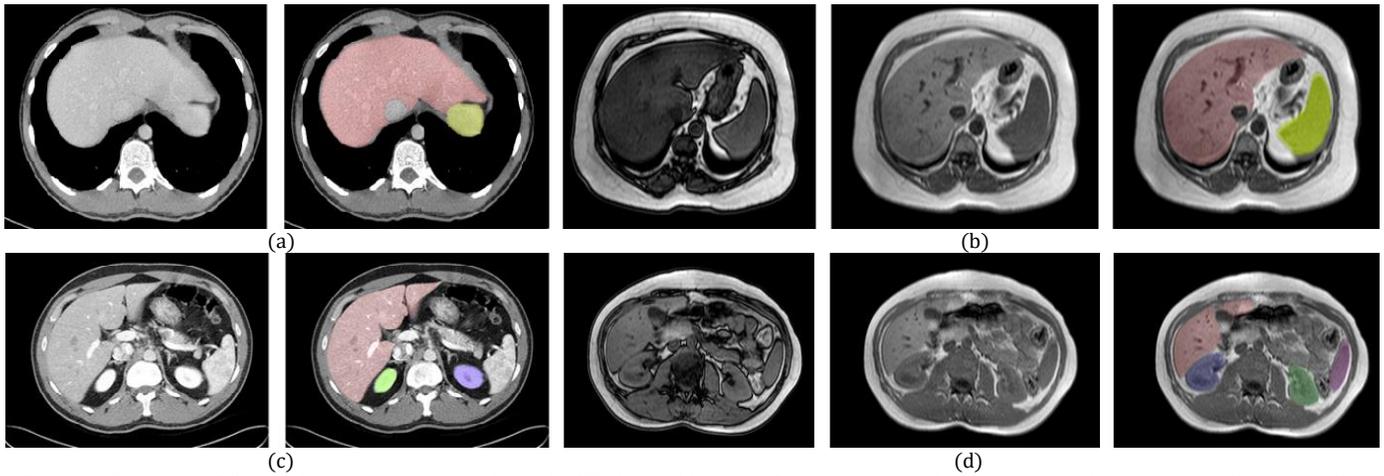

Fig. 4. (a) Examples of liver and spleen parenchyma tissues in CT. Original image (left) and segmented image (right). Red represents the liver and yellow represents the spleen. (b) Examples of liver and spleen parenchyma tissues in MR. Out-phase image (left), in-phase image (middle) and segmented image (right). Red represents liver and yellow represent spleen. (c) Examples of liver and kidneys in CT. Original image (left) and segmented image (right). Red represents the liver and green/blue represents the right/left kidneys. (d) Examples of liver and kidneys in MR. Out-phase image (left), in-phase image (middle) and segmented image (right). Red represents the liver and green/blue represents the right/left kidneys.

both of which contain upper abdomen image series of 20 patients (i.e., 40 individuals in total). 8 patients in the dataset is utilized in the test while 12 of them is used in the training step. These patients are potential liver donors having a healthy liver (i.e. no focal lesions such as tumors or no evidence of diffuse parenchymal diseases). The CT images are obtained from contrast enhanced CT examination at portal venous phase in which adequate parenchymal enhancement is provided for segmentation. In this phase, the parenchyma is also affected due to the blood supply by the portal vein creating a challenging non-homogeneous intensity distribution throughout the liver. For all datasets, patient orientation and alignment are the same.

The MR dataset covers dual sequence (axial GRE T1 weighted in-phase and out of phase) MRI images obtained from a 1.5T Philips MRI scanner. Dual sequence MRI imaging is a fat suppression sequence, which uses the time differences in the z-axis recoveries of fat and water protons. The signal is acquired twice: first when water and fat protons are in phase; and second, when they are out of phase (while excited protons are returning to their first position). Images in these two phases are registered. Visual samples for CT and MRI datasets are illustrated in Fig. 3.

CT and MRI images have significantly diverse characteristics because of the differences in the underlying working principles of the two modalities. In single modality segmentation systems, these differences are omitted. However, for cross-domain applications, they have critical importance. CT and MRI introduce their own advantages and disadvantages for liver imaging. One of the most important differences is the acquisition times. A CT scan of the upper abdominal area usually takes less than a minute, while an MRI scan can take up to 20 minutes. This difference makes CT a useful option for contrast-enhanced imaging, as the contrast agent remains in the hepatic veins for only 20 to 30 seconds. Moreover, MRI acquisition is subject to patient movements more due to the long acquisition time. This makes CT images more preferable for measuring liver volume as the organ border appears sharper and clearer compared to MRI especially for the patients having breath-holding problems.

On the other hand, MRI can produce a much better contrast resolution between soft tissues, making it ideal for the detailed analysis of focal lesions and for determining the boundaries of the organ at low gradient boundaries. For instance, in CT images, the density and position of the pixels/voxels corresponding to liver boundaries can be very close to the ones that belong to the spleen, gastric wall, muscle tissues or even the right kidney. The distinction between these boundary pixels/voxels can further diminish due to the partial volume effect. Some of the problematic cases for liver segmentation does not exist in MRI images, which can provide much higher contrast resolution for soft tissues. Visual examples for the case are shown in Fig. 4.

Another drawback of MRI against CT is the inter-slice distance. A typical MRI scan procedure in the routine clinical workflow for the upper abdomen area has much higher inter-slice distance, which creates a sparse representation of organs and sudden changes in their sizes and shapes. Despite the well-established Hounsfield units of the liver in CT, un-calibrated intensity ranges might change the appearance of the liver during MRI acquisitions. Several factors might cause these deviations in intensity of the acquired signals and results with significant changes about the visibility of the same tissue inside a dataset. Furthermore, due to the nature of MRI imaging, the same organ might appear in a different intensity range for different datasets even for the same pulse sequences. However, the behavior of the histogram in realistic MRI data has considerable intensity fluctuations.

*B. Utilized Metrics and Evaluation Strategy*

Various metrics are presented in the literature to compare two 2D and 3D objects (segmentation outcome) in terms of their visual similarities. However, since each metric examines the similarities from different perspectives, none of these metrics is sufficient to evaluate the scoring alone. For unpaired CT

TABLE 2. PERFORMANCE OF CROSS-DOMAIN SEGMENTATION METHODS. RED BOLD INDICATES THE BEST PERFORMANCE FOR EACH METRIC.

| Methods | Performance Metrics (CT / MR) | | | | | |
|---|---|---|---|---|---|---|
| | VO | RVD | ASSD | RMSD | MSSD | Overall |
| Cond-DCN [17] | 86.42/81.26 | 80.86/26.69 | 87.02/91.46 | 82.88/91.04 | 22.79/76.21 | 71.99/73.33 – 72.66 |
| Att-DCN [18] | 89.35/83.32 | 84.26/40.60 | 89.42/92.64 | 86.60/92.96 | 60.09/86.07 | 81.94/79.12 – 80.53 |
| Pre-BN-DCN w/o CAN | 89.46/84.29 | 85.98/23.35 | 88.34/95.82 | 85.17/94.87 | 62.13/87.04 | 82.22/77.07 – 79.64 |
| Post-BN-DCN w/o CAN | 91.06/87.66 | 79.22/41.43 | 88.25/96.62 | 83.62/95.67 | 52.58/88.73 | 79.01/82.02 – 80.51 |
| Post-BN-DCN | **91.37**/**87.46** | 82.70/46.97 | 88.91/**96.63** | 84.10/**95.74** | 54.37/**89.49** | 80.29/**83.26** – 81.75 |
| Pre-IN-DCN (ours) | 90.93/84.77 | 88.35/45.02 | 91.33/93.00 | 89.08/93.46 | **65.45**/88.51 | **85.03**/80.95 – 82.99 |
| Pre-BN-DCN (ours) | 91.30/84.68 | **89.72**/**56.87** | **92.03**/93.09 | **89.92**/93.25 | 58.01/86.17 | 84.20/82.81 – **83.50** |

and MR liver dataset, the following metrics are used and the final score is calculated by the average of these five metrics:
- Volumetric overlap (VO)
- Relative volume difference (RVD)
- Average symmetric surface distance (ASSD)
- RMS symmetric surface distance (RMSD)
- Maximum symmetric surface distance (MSSD)

For Covid-19 dataset, pixel precision (PR) and Volumetric overlap (VO) are utilized. Each metric is briefly summarized in this section.

**Volumetric overlap (VO):** In this metric, the number of voxels at the intersection of the segmentation result $V_{seg}$ and the reference result $V_{ref}$ is divided by the number of voxels in the union of segmentation and reference results:

$$VO = \frac{V_{seg} \cap V_{ref}}{V_{seg} \cup V_{ref}} \; x \; 100 \quad (7)$$

The value of volumetric error rate is 100 for flawless segmentation. In the worst case, where there is no overlap between segmentation and reference, it produces zero.

**Relative volume difference (RVD):** The total volume difference between segmentation result and reference image is divided by the total volume of the reference image. If a perfect segmentation is performed, the output of RVD is zero. For segmentation results that are less successful, it is greater than zero:

$$RVD = \left| \frac{V_{seg} - V_{ref}}{V_{ref}} \right| x100 \quad (8)$$

Since the lower relative volume difference represents higher performance, the value of the metric is converted to the score by using the inverse ratio as shown in Table 1.

**Average symmetric surface distance (ASSD):** Symmetrical surface distance (SSD) metrics are beneficial for measuring the similarity of two objects in terms of their shapes. In the SSD method, two three-dimensional objects are aligned so that their centers overlap. Then all the Euclidian distances between a boundary voxel in the first object and all voxels of the boundary of the second object are calculated. The lowest of the calculated distances is determined as the SSD of the voxel in the first object from the second object $(d(x, A))$. This process is repeated for all border voxels in both objects:

$$d(x, A) = \min_{y \in A}(d(x, y)) \quad (9)$$

ASSD is the arithmetic mean of the measured symmetrical surface distances of the two objects and is used in particular to evaluate the volume differences:

$$ASSD = \frac{1}{|V_{seg}|+|V_{ref}|} \times \left( \sum_{x \in V_{seg}} d(x, V_{ref}) + \sum_{y \in V_{ref}} d(y, V_{seg}) \right) \quad (10)$$

This metric gives zero if two 3D objects are the same. There is no upper limit for non-similar segmentations.

**RMS symmetric surface distance (RMSD):** RMSD is calculated by the Root Mean Square (RMS) of the distance measurements of the two objects:

$$RMSSSD = \sqrt{\frac{1}{|V_{seg}|+|V_{ref}|}} \times \sqrt{\sum_{x \in V_{seg}} d^2(x, V_{ref}) + \sum_{y \in V_{ref}} d^2(y, V_{seg})} \quad (11)$$

**Maximum symmetric surface distance (MSSD):** MSSD represents the largest difference between the two objects. This metric has crucial importance because it is particularly useful in evaluating applications where surgical precision is critical.

$$MSSD = \max \left( \min_{x,y \in V_{ref}} (d(x, y)) \right) \quad (12)$$

*C. Baseline Methods*

To compare the performance of the proposed method, two DCN architectures are selected as baselines. In short, Cond-DCN [17] classifies the modalities of data with an auxiliary network architecture. Later, these prediction outputs are conditioned in the semantic segmentation with each DCN layer. The key aspect of this model is to independently learn different solutions for each modality. Note that in the training phase of the model, labels related to domains are made available to the network, even if it is not practical in real scenario.

In addition, Att-DCN [18] is used to improve the selectivity of skip connections by taking advantage of the idea of attention networks. More specifically, it uses the responses of decoder and encoder modules for the same level as the input to calculate attention heat-maps and select the best solutions. By this way, a subset of responses from encoder layers is used in the segmentation step.

*D. Experimental Results on Upaired CT and MR Dataset*

In the experiments, we first compare our results with the baseline techniques in the literature. The experimental results are illustrated in Table 2. In addition, the last column is reserved for average scores of CT and MRI performance for each method. Note that $T$ value that indicates the number of input slices is set to 1.

TABLE 3. PERCENTAGE OF RESPONSES FORWARDING TO THE NEXT LAYERS. NOTE THAT THIS FEATURE SHOWS THE SPARSITY OF KERNELS.

| Methods | Modality of Data | |
|---|---|---|
| | CT | MR |
| **Pre-BN-DCN w/o CAN** | 0.596 | 0.627 |
| **Post-BN-DCN w/o CAN** | 0.584 | 0.592 |
| **Post-BN-DCN** | 0.491 | 0.480 |
| **Pre-BN-DCN (ours)** | 0.277 | 0.411 |

From these results, the proposed structure outperforms all baseline methods with a large margin. This shows the generalization capacity of our method, especially for cross-domain learning problem. In particular, Cond-DCN [17], obtains worse performance even if an auxiliary network that estimates modality labels is used.

Parameter overfitting with limited data can be one of leading factors since a different parameter set needs to be learned from data and it multiplies the number of trainable parameters in the problem. In the literature, as mentioned, the studies show that multi-task learning is suitable to eliminate this problem [36].

As indicated, use of pre-structure (ours) alone yields slightly worse performance than post-structure if no adversarial loss is applied. Strictly speaking, parameter underfitting as a result of loss of sparsity can be shown as the main reason for the results. For this purpose, we integrate an adversarial loss to have sparsity and regularizes the network parameters especially for grouping data information in clustering fashion. By this way, misclassified parts of segments (i.e. false positive) are penalized with adversarial loss. Notice that improvement compared to pre-structure is high compared post-structure. Reducing the effect of trainable parameters to normalization layers with pre-structure is positively affected the adversarial loss to group more domain-specific statistics.

To prove the statement empirically, Table 3 shows the percentage of responses that are active in next layers for various DCN structures. It is clear that pre-structure with adversarial loss yields the best sparsity (i.e. ideally lower is better since more confident prediction can be made) to the responses.

Moreover, either use of either Batch normalization (BN-DCN) or Instance normalization (IN-DCN) cannot outperform the other one but obtains better compared to baselines. Note that this shows that our solution does not overfit to a particular architecture.

Last but not least, we compare the performance of the proposed method (i.e., the effect of cross-domain learning with a single architecture) with individually trained models on each domain. In Table 4, the proposed cross-domain model significantly improves performance compared to these baselines. This experiment validates the impact of the method for the cross-domain learning.

TABLE 4. PERFORMANCE FOR INDIVIDUAL MODALITIES AND CROSS-DOMAIN.

| Methods | Modality of Data | |
|---|---|---|
| | CT | MR |
| **DCN for CT** | 75.81 | - |
| **DCN for MR** | - | 79.47 |
| **Pre-BN-DCN (ours)** | **84.20** | **82.81** |

For the second part of the experiment, the impact of slice number is investigated as an ablation study. The results are given in Table 5. Similarly, last columns indicate the average scores. From the results, batch normalization with $T=0$ and instance normalization with $T=1$ obtain the best performance compared to the others. When we analyze the data visually, we observe that MR data is more sensitive to noise due to the undesired movements during long acquisition time and their low spatial resolutions. Therefore, instance normalization obtains better results for MR while batch normalization outperforms for CT in most cases. This observation is another critical result which can lead the future studies.

Increasing slice numbers does not improve the performance of semantic segmentation (i.e., more information is considered in the feature extraction step). The main reason is that the inter-slice distances for each modality are different (i.e., approximate resolution of CT is two times higher than MR data). Therefore, smaller slice numbers may be the best option to realize a solution for cross-domain scenario with a single model unless an extra registration step is added.

### E. Experimental Results on Covid-19 dataset

In the experiments, we only report the post and pre-structure scores for the dataset. The results are illustrated in Table 6. From the results, it can be seen that pre-structure predominantly achieves the best performance for all lung parts. In particular, the performance gap in the Covid part is positively large for the proposed model. The main reason is that, as stated, even if all dataset is in CT format, due to the variations in sensor type/noise level etc., each scanned data acts as in cross-domain cases. Thus, lack of a sufficient amount of data for the learning process becomes problematic for the dataset. However, the proposed method is robust and computes relevant features for all data type.

Another important observation is that false prediction rate is high for post-structure even if adversarial loss is utilized. Similarly, this can be explained with the confidence of the predictions with the proposed structure.

TABLE. 5. PERFORMANCE OF CROSS-DOMAIN FOR DIFFERENT $T$ VALUES. RED BOLD INDICATES THE BEST PERFORMANCE FOR EACH METRIC.

| Methods | Performance Metrics (CT / MR) | | | | | |
|---|---|---|---|---|---|---|
| | VO | RVD | ASSD | RMSD | MSSD | Overall |
| **IN-DCN (T=0)** | 89.56/**86.15** | 77.04/65.11 | 89.07/**93.28** | 85.33/**93.67** | 48.10/87.47 | 77.82/**85.13** – 81.47 |
| **IN-DCN (T=1)** | 90.93/84.77 | 88.35/45.02 | 91.33/93.00 | 89.08/93.46 | **65.45**/**88.51** | **85.03**/80.95 – 82.99 |
| **IN-DCN (T=2)** | 88.69/83.94 | 78.40/50.09 | 89.29/93.03 | 85.90/93.17 | 49.27/86.28 | 78.31/81.30 – 79.80 |
| **IN-DCN (T=3)** | 89.69/81.59 | 88.05/45.52 | 90.84/92.43 | 88.06/92.25 | 54.28/83.73 | 82.22/79.11 – 80.66 |
| **BN-DCN (T=0)** | **91.63**/85.03 | 87.92/**66.74** | 91.68/92.75 | 89.41/92.81 | 61.44/86.24 | 84.42/84.70 – **84.56** |
| **BN-DCN (T=1)** | 91.30/84.68 | **89.72**/56.87 | **92.03**/93.09 | **89.92**/93.25 | 58.01/86.17 | 84.20/82.81 – 83.50 |
| **BN-DCN (T=2)** | 90.53/83.58 | 86.45/50.19 | 91.34/93.02 | 88.83/93.30 | 57.32/86.55 | 82.89/81.33 – 82.11 |
| **BN-DCN (T=3)** | 89.29/81.61 | 85.39/35.83 | 90.62/92.52 | 87.75/92.62 | 54.65/85.79 | 81.54/77.67 – 79.60 |

TABLE 6. PRECISION AND IOU RESULTS ON COVID-19 DATASET. RED BOLD INDICATES THE BEST PERFORMANCE FOR EACH METRIC

| Methods | Parts (PR / IoU) | | |
|---|---|---|---|
| | Lung | Covid | BG |
| Post-BN-DCN | **96.47** / 38.89 | 30.38 / 17.91 | 89.74 / - |
| Pre-BN-DCN (ours) | 95.48 / **41.13** | **40.39** / **22.16** | **92.84** / - |

## V. CONCLUSION

In this paper, we propose a novel DCN model for semantic segmentation problem for biomedical imaging. The superiority of the model is that it computes segmentation predictions from different modalities without knowing priors of data in training and inference steps. First, we show that conventional models do not work well on cross-domain problem since this solution saturates to compute singly shared features from data, thus the conventional structure needs to be revised based on the observations explained in the manuscript. Reordering normalization and convolution operators allows us to improve performance by enabling more statistical information past to the network. However, use of this step alone underestimate the solution so that adversarial loss is integrated to the model to regularize the network and help to cluster representations in an unsupervised settings.

The experimental results confirm that the proposed model for CT-MR liver segmentation achieves state-of-the-art performance compared to all baseline methods. Moreover, intra-class modality differences can be captured precisely with the proposed method as validated in the experiments of Covid-19 dataset.

ACKNOWLEDGMENT

The authors gratefully acknowledge the support of NVIDIA Corporation with the donation of GPUs used for this research.

REFERENCES

[1] A. Krizhevsky, I. Sutskever, and G. E. Hinton, "Imagenet classification with deep convolutional neural networks," in *Proceedings of the 25th International Conference on Neural Information Processing Systems*, 2012, pp. 1097–1105.
[2] K. He, X. Zhang, S. Ren, and J. Sun, "Deep residual learning for image recognition," in *In Proceedings of the IEEE conference on computer vision and pattern recognition,* 2016, pp. 770-778.
[3] E. A. Rios Piedra, B. M. Ellingson, R. K. Taira, S. El-Saden, A. A. T. Bui, and W. Hsu, "Brain tumor segmentation by variability characterization of tumor boundaries," In *International Workshop on Brainlesion: Glioma, Multiple Sclerosis, Stroke and Traumatic Brain Injuries*, 2016, pp. 206–216.
[4] R. Ayachi and N. Ben Amor, "Brain tumor segmentation using support vector machines," In *European Conference on Symbolic and Quantitative Approaches to Reasoning and Uncertainty*, 2009, pp. 736–747.
[5] E Shelhamer, J. Long, and T. Darrell, "Fully convolutional networks for semantic segmentation," In *Proceedings of the IEEE conference on computer vision and pattern recognition*, 2015, pp. 3431-3440.
[6] O. Ronneberger, P. Fischer, and T. Brox, "U-net: Convolutional networks for biomedical image segmentation," In *International Conference on Medical image computing and computer-assisted intervention*, 2015, pp. 234-241.
[7] K. Kamnitsas, C. Ledig, V. F. J. Newcombe, J. P. Simpson, A. D. Kane, D. K. Menon, et al., "Efficient multi-scale 3d CNN with fully connected CRF for accurate brain lesion segmentation," *Medical image analysis*, pp. 61-78, 2017.
[8] S. Ioffe and C. Szegedy, "Batch normalization: Accelerating deep network training by reducing internal covariate shift," *ArXiv*:1502.03167, 2015.
[9] Ulyanov, Dmitry, Andrea Vedaldi, and Victor Lempitsky. "Instance normalization: The missing ingredient for fast stylization." *ArXiv*:1607.08022, 2016.
[10] T. Salimans, and D. P. Kingma, "Weight normalization: A simple reparameterization to accelerate training of deep neural networks," in *Advances in Neural Information Processing Systems*, 2016, pp. 901-909.
[11] K. He, X. Zhang, S. Ren, and J. Sun, "Delving deep into rectifiers: Surpassing human-level performance on imagenet classification," In *Proceedings of the IEEE international conference on computer vision*, 2015, pp. 1026-1034.
[12] I. Goodfellow, J. Pouget-Abadie, M. Mirza, B. Xu, D. Warde-Farley, S. Ozair, et al., "Generative adversarial nets," in *Advances in Neural Information Processing Systems*, 2014, pp. 2672–2680.
[13] M. Rezaei, K. Harmuth, W. Gierke, T. Kellermeier, M. Fischer, H. Yang, et al., "Conditional adversarial network for semantic segmentation of brain tumor," in *International MICCAI Brainlesion Workshop*, 2017, pp. 241-252.
[14] P. Luc, C. Couprie, S. Chintala, and J. Verbeek, "Semantic segmentation using adversarial networks," *ArXiv*:1611.08408, 2016.
[15] P. Moeskops, M. Veta, M. W. Lafarge, K. A. J. Eppenhof, and J. P. W. Pluim, "Adversarial training and dilated convolutions for brain MRI segmentation," *ArXiv*:1707.03195, 2017.
[16] D. P. Kingma and J. Ba, "Adam: A method for stochastic optimization," *ArXiv*:1412.6980, 2014.
[17] V.V. Valindria, N. Pawlowski, M. Rajchl, I. Lavdas, E. O. Aboagye, A. G. Rockall, et al., "Multi-modal learning from unpaired images: Application to multi-organ segmentation in ct and mri," in *IEEE Winter Conference on Applications of Computer Vision*, 2018, pp. 547-556.
[18] O. Oktay, J. Schlemper, L.L. Folgoc, M. Lee, M. Heinrich, K. Misawa, et al., "Attention U-Net: learning where to look for the pancreas," *ArXiv*:1804.03999, 2018.
[19] I. Goodfellow, J. Shlens, and C. Szegedy, "Explaining and harnessing adversarial examples," *ArXiv*:1804.039991412.6572, 2014.
[20] I. Lavdas, B. Glocker, K. Kamnitsas, D. Rueckert, H. Mair, A. Sandhu, et al., "Fully automatic, multi-organ segmentation in normal whole body magnetic resonance imaging (mri), using classification forests (cfs), convolutional neural networks (cnns) and a multi-atlas (ma) approach," *Medical Physics*, pp. 5210-5220, 2017.
[21] M. Havaei, A. Davy, D. Warde-Farley, A. Biard, A. Courville, Y. Bengio, et al., "Brain tumor segmentation with deep neural networks". *Medical image analysis*, pp. 18–31, 2017.
[22] K. Kamnitsas, C. Ledig, V. F. Newcombe, J. P. Simpson, A. D. Kane, D. K. Menon, et al., "Efficient multi-scale 3d cnn with fully connected crf for accurate brain lesion segmentation." *Medical image analysis*, pp. 61–78, 2017.
[23] S.A. Rebuffi, H. Bilen, and A. Vedaldi. "Learning multiple visual domains with residual adapters." In Advances in Neural Information Processing Systems, pp. 506-516. 2017.
[24] H. Bilen, and Andrea Vedaldi. "Universal representations: The missing link between faces, text, planktons, and cat breeds." arXiv preprint arXiv:1701.07275 (2017).
[25] O. Hadad, R. Bakalo, R. Ben-Ari, S. Hashoul, and G. Amit, "Classification of breast lesions using cross-modal deep learning". In International Symposium on Biomedical Imaging, 2017, pp. 109-112.
[26] X. Tang, B. Du, J. Huang, Z. Wang, and L. Zhang, "On combining active and transfer learning for medical data classification," *IET Computer Vision*, pp. 194-205, 2018.
[27] A. A. Taha and A. Hanbury, "Metrics for evaluating 3D medical image segmentation: Analysis, selection, and tool," in *BMC medical imaging*, pp. 15-29, 2015.
[28] T. Heimann, B. V. Ginneken, and M.A. Styner, "Comparison and Evaluation of Methods for Liver Segmentation From CT Datasets," *IEEE Transactions on Medical Imaging*, pp. 1251–1265, 2009.
[29] C. Ouyang, K. Kamnitsas, C. Biffi, J. Duan, and D. Rueckert. "Data Efficient Unsupervised Domain Adaptation for Cross-Modality Image Segmentation," *ArXiv*:1907.02766, 2019.
[30] Y. Xue, T. Xu, H. Zhang, L. R. Long, and X. Huang, "Segan: Adversarial network with multi-scale l 1 loss for medical image segmentation," *Neuroinformatics,* pp. 383-392, 2018.
[31] Y. Zhan, M. Dewan and X.S. Zhou, "Cross modality deformable segmentation using hierarchical clustering and learning", International

Conference on Medical Image Computing and Computer-Assisted Intervention, 2009

[32] S.A. Sriram, A. Paul, Y. Zhu, V. Sandfort, P.J. Pickhardt and R. M. Summers, "Multilevel UNet for pancreas segmentation from non-contrast CT scans through domain adaptation", Medical Imaging 2020: Computer-Aided Diagnosis, 2020.

[33] Q. Dou, C. Ouyang, C. Chen, H. Chen and P.A. Heng, "Unsupervised cross-modality domain adaptation of convnets for biomedical image segmentations with adversarial loss", arXiv preprint arXiv:1804.10916, 2018.

[34] M Arjovsky, S. Chintala and L. Bottou, "Wasserstein generative adversarial networks", International Conference on Machine Learning, 2017.

[35] F. Milletari, N. Navab and S.A. Ahmadi, "V-Net: Fully Convolutional Neural Networks for Volumetric Medical Image Segmentation", International Conference on 3D Vision, 2016.

[36] S. Liu, E. Johns and J. A. Davison, "End-to-end multi-task learning with attention", IEEE Conference on Computer Vision and Pattern Recognition, 2019.

[37] S. Mukherjee, H. Asnani, E. Lin, and S. Kannan, "Clustergan: Latent space clustering in generative adversarial networks." In Proceedings of the AAAI Conference on Artificial Intelligence, vol. 33, pp. 4610-4617. 2019.

[38] X. Yang, C. Deng, F. Zheng, J. Yan, and W. Liu. "Deep spectral clustering using dual autoencoder network." In Proceedings of the IEEE Conference on Computer Vision and Pattern Recognition, pp. 4066-4075. 2019.